\documentclass[fleqn,twoside]{article}
\usepackage{espcrc2}

\usepackage{graphicx}
\usepackage[figuresright]{rotating}

\newcommand{\AmS}{{\protect\the\textfont2
  A\kern-.1667em\lower.5ex\hbox{M}\kern-.125emS}}

\title{Collider implications of models with extra dimensions}

\author{C. Macesanu\address[OSU]{Department of Physics, Oklahoma State 
        University, \\ 
        Stillwater, Oklahoma, 78074, USA}, 
        C.D. McMullen\address{Department of Physics, 
        Penn State Altoona University,\\
        Altoona, PA, 16601, USA}
        and
        S. Nandi\addressmark[OSU]\thanks{Invited talk presented by S. Nandi
        at the ICHEP 2002, Amsterdam, The Netherlands, July 24-31, 2002}
        }
\begin{document}

\begin{abstract}
 We discuss the collider signals of large extra dimensions in which gravity
as well as the SM particles propagate into the extra dimensions. These
signals arise either from the production of Kaluza-Klein excitations of the
SM particles and their subsequent decay, or from their off-shell exchanges. 
Depending on the scenario, the dominant signals are two high $p_T$ jets +
missing energy, two high $p_T$ photons + missing energy and soft leptons, 
or a combination of photon + jet and missing energy. For the scenario in
which only the gauge bosons propagate into the extra dimensions, 
Tevatron Run II (LHC) can observe such signals up to a compactification scale 
of about 2 TeV (7 TeV), while for the case of universal extra dimensions, 
the corresponding limits are about 600 GeV (3 TeV) respectively.
\vspace{1pc}
\end{abstract}

\maketitle

\section{Introduction}
Recent developements \cite{string} in string theory have led to the construction
of models in which the compactification scale of the extra dimensions can 
be as large as inverse TeV \cite{led} or even submilimeter \cite{mil}. In this
later case, due to experimental constraints, only gravity can be allowed 
to propagate into the extra dimensions, while for the inverse TeV scale,
both gravity and the Standard Model particles can be in the bulk.
The phenomenological implications for the models with submilimeter extra
dimensions have been extensively discussed in the past few years.  In 
this talk, we will concentrate on the implications of the inverse TeV scale
extra dimensions. In this case, there are two major possible scenarios:
only  the SM gauge bosons propagate into the extra dimensions, or all
the SM particles propagate in  the bulk (also called the Universal Extra 
Dimensions scenario). A hybrid scenario \cite{ln} 
in which the model accomodates 
sub-mm (in which only gravity propagates) as well as inverse TeV scale 
extra dimensions (where SM particles also propagate) has additional interesting
features. Such a model gives rise to deviations from Newton's law
of gravity at sub-mm distances and several interesting astrophysical phenomena. 
 It also leads to new physics signals at high energy colliders 
due to the production of KK excitations of SM particles.

\section{Only gauge bosons in the bulk}
In this scenario, only the SM gauge bosons propagate into the extra dimensions.
The SM fermions are confined to the usual 3+1 dimensional wall, the
$D_3$-brane. We will consider the case of only one extra dimension of
inverse TeV size. The collider signals will be similar for more than
one extra dimension. The 5th dimension is compactified on an $S^1/Z_2$ 
orbifold (with $y \rightarrow -y$ symmetry) and the physical space is 
$M_4 \times S^1/Z_2$, where $M_4$ is the Minkowski space. The five dimensional
gauge fields can be expanded as:
\begin{eqnarray}\label{gauge} A_{\mu}^a  & = &
\frac{1}{\sqrt{\pi R}}\left[ A_{\mu, 0}^a  +
\sqrt{2} \sum_{n=1}^{\infty}A_{\mu ,n}^{a}  \cos(\frac{n y}{R}) \right] 
\nonumber \\
A_{4}^a  & = & \frac{\sqrt{2}}{\sqrt{\pi R}}
\sum_{n=1}^{\infty}A_{4,n}^{a}  \sin(\frac{n y}{R}) \, .
\end{eqnarray}
Here $A_{\mu ,n}^{a}$ are the $n$th KK excitations of the gauge bosons
with masses $m_n \equiv n/R$. The collider signals will mainly come from 
the production of the KK excitations of the gluons ($g_n$) and their
subsequent decays, or through the off-shell exchange of  the KK gluons
in the usual SM processes. In this model, we can have new couplings
such as $\bar{q} q g_n, g g_n g_n, g_n g_m g_p$ etc. (The detailed 
interactions and the Feynman rules for all the allowed couplings can 
be found in \cite{cnd}). The new collider signals arise from the processes
$q \bar{q} \rightarrow g^* \rightarrow q \bar{q}$ (dijet signal);
$q \bar{q} \rightarrow g^* g \rightarrow q \bar{q} g$ (3-jet signal) and
$q \bar{q} \rightarrow g^* g^* \rightarrow (q \bar{q}) (q \bar{q})$ 
(4-jet signal). (Here $g^*$ stands for a general KK excitation of the gluon).
Since these jets are coming from the decay of a very heavy particle,
they carry much higher $p_T$ than the jets produced by the usual SM processes.
The high $p_T$ dijet signal is observable over the usual QCD background,
and gives the reach for the compactification scale of about 2 TeV at the 
Tevatron Run II and 7 TeV at the LHC \cite{cnd}.

\section{Universal extra dimensions}

In this scenario, all SM particles propagate in the bulk \cite{ACD}. Again, for
simplicity, we consider the case of only one extra dimension of
inverse TeV size. With compactifiation on an $S^1/Z_2$ orbifold,
the decomposition of the 5-dimensional gauge fields is the same as
above (Eq. (\ref{gauge})). In 5 dimensions, the fermion fields are vectorlike;
in order to obtain the chiral fermions of the SM, there are needed
two sets of fermionic fields, with different properties under the $Z_2$ 
parity. Thus, we introduce an SU(2) doublet, whose left-handed fields
are even and right-handed fields are odd under $Z_2$ parity:
\begin{eqnarray} Q  &  = &
\frac{1}{\sqrt{\pi R}} \left\{ Q_{L} + \sqrt{2} \sum_{n=1}^{\infty}
\left[ Q_L^n  \cos \left(\frac{n y}{R} \right) \right. \right.  \nonumber \\
& & \left. \left. + Q_R^n  \sin \left(\frac{n y}{R} \right) \right] \right\}
\end{eqnarray}
and an SU(2) singlet with opposite properties:
\begin{eqnarray} q  &  = &
\frac{1}{\sqrt{\pi R}} \left\{ q_{R} + \sqrt{2} \sum_{n=1}^{\infty}
\left[ q_R^n  \cos \left(\frac{n y}{R} \right) \right. \right.  \nonumber \\
& & \left. \left. + q_L^n  \sin \left(\frac{n y}{R} \right) \right] \right\}.
\end{eqnarray}
With these definitions, we obtain the correct chiral nature of the SM fermions
(which are build from the zero modes of the above fields). For the KK 
exciations we then obtain two towers for each SM fermionic field:
$q_n^{\bullet} = Q_L^n + Q_R^n$ and $q_n^{\circ} = q_L^n + q_R^n$ (ignoring
small mixing due to Yukawa interactions). The masses of the fermionic
excitations are given by $m_n^2 = (n/R)^2 + m_{SM}^2$. The Feynman rules 
for the allowed couplings in this model can be found in \cite{cmn}.
The most interesting
feature of this model is the KK number conservation at tree level. 
As a consequence,
excitations of gluons and quarks can only be pair produced at colliders, which
results in significantly lower limits (around 350 GeV) set on the 
compactification  scale $1/R$ from current experiments \cite{ACD,cmn}. 

\subsection{Decay mechanisms}

There are two possible decay scenarios for KK excitations in the UED models.\\
(i) \underline{Loop effects:} In first order, the masses of the KK
excitations of the gluons and light quarks are essentialy degenerate
at each level. However, loop corrections due to virtual particles propagating
in the bulk, as well as boundary terms due to the breaking of
translational invariance at the orbifold boundaries
 will remove this degeneracy. 
With some assumptions (which include universality for the boundary terms as
well as the introduction of a cutoff scale $\Lambda$) it is possible to 
compute the corrections to the KK masses at all levels
 \cite{schmaltz}. Taking $\Lambda = 10 $ TeV, the relative 
corrections
to the first KK level masses are as follows: $\delta m_{g^*} \simeq 30\%$,
$\delta m_{q^*} \simeq 20\%$ and $\delta m_{W^*,Z^*,l^*} \le 10\%$.
The lightest KK particle (LKP) will then be the $\gamma^*$,
 and, due to KK number conservation, it will be stable, with all other
first level excitations decaying to it. In this scenario, the decay modes
for the quark and gluon excitations produced at a hadron collider will be:
$$
g_1 \rightarrow q \bar{q_1^{\bullet}}, q \bar{q_1^{\circ}} \ ; \ \
q_1^{\circ} \rightarrow q \gamma^* \ ; \ \hbox{and} 
$$
$$q^{\bullet}_1 \rightarrow q\ Z^*_1 \rightarrow 
q\ l\ l^{\bullet}_1 \rightarrow q\  l\ l\ \gamma^* \ ,
  \hbox{Br.} \sim 33\%, \ \hbox{or}
$$
\begin{equation}\label{cascade}
q^{\bullet}_1 \rightarrow q\ W^*_1 \rightarrow 
q\ l' \ l^{\bullet}_1
 \rightarrow q\  l'\ l \ \gamma^* \ ,
  \hbox{Br.} \sim 65\%
\end{equation}
The final states in this scenario contains two $\gamma^*$, plus
soft jets and leptons. If this is the only decay mode, then the observable
signal will be soft leptons plus large missing energy (from the LKP). Then,
the Tevatron Run II reach is about 300 GeV, while the LHC reach is about 1.5 TeV
\cite{schmaltz2}.  

\noindent
(ii)\underline{Gravity mediated decays} (fat brane scenario): In this scenario,
gravity-mediated interactions break KK number conservation, allowing 
decays such as 
\begin{equation}\label{grav}
g^* \rightarrow g G_n, \  q^* \rightarrow q G_n, \
\gamma^* \rightarrow \gamma G_n
\end{equation}
where $G_n$ is the KK excitation of
the graviton. We assume here that gravity propagates in $N$ extra dimensions
of submilimeter size (such that the above decays are kinematically allowed),
while the SM particles can propagate only a small distance -inverse TeV size-
along one of these dimensions, i.e., the SM matter lives on a fat 
$D_3$ brane \cite{rujula}. In this case, one obtains a form factor describing 
the superposition of the matter wave function with the graviton wave function 
along the fifth dimension, which gives rise to the couplings which allow
the decays in Eq. (\ref{grav}).

\subsection{Collider signals for the fat brane scenario}
In UED models, KK excitations of quarks and gluon are pair produced at 
hadron colliders. The final state signal will depend on which of the 
above decay mechanism, (i) or (ii), dominates. If $\Gamma_{(ii)}\gg
\Gamma_{(i)}$, then the final state signal will be two jets plus
missing energy due to the escaping graviton excitations. The transverse
momenta of the observable jets will be quite large, as will be the
missing transverse energy (for an analysis, see \cite{cmn}). This will
allow elimination of the SM backgrounds with suitable $p_T$ and $\not{E_T}$
cuts. The Tevatron Run II reach in this mode is about 500 GeV, while
the LHC reach is about 3 TeV \cite{cmn}.  

If the decay mechanism (i) dominates over (ii), i.e. $\Gamma_{(i)}\gg
\Gamma_{(ii)}$, then the KK excitations produced at colliders will first
cascade decay to the LKP as in (\ref{cascade}). However, due to gravitational
interaction the LKP is not stable anymore and it will decay: 
$\gamma^* \rightarrow \gamma G_n$. The signal in this case will the be 
two hard photons plus large $\not{E_T}$ (and some soft leptons and jets).
 Collider reach in this case is around 550 GeV at the Tevatron
Run II and 3 TeV at the LHC \cite{cmn2}.

If we are in a region of the parameter space where $\Gamma_{(i)} \simeq
\Gamma_{(ii)}$, then the signal will be a mixture of the modes discussed
above, that is $\gamma \gamma, \gamma + $jet, 2 jets, plus large $\not{E_T}$
in the final states. For details, see \cite{cmn2}.

\section{Conclusions}

Models with extra dimensions in the inverse TeV range can lead to
interesting phenomenology at high energy colliders. Depending on the
scenario, the signals may be high $p_T$ dijets plus 
missing ${E_T}$, or
even high $p_T$ diphotons plus $\not{E_T}$. 
Tevatron reach is from $\sim$ 400 GeV
to 2 TeV, while LHC reach is from $\sim$ 1.5 TeV to around 7 TeV, depending
on the model (the
higher limits are for the case when only the SM gauge bosons 
propagate in the extra dimensions). For low compactification scales, 
LHC may observe the second level KK excitations, which will be a clear signal
for the existence of extra dimensions.

\vspace{0.3cm}
\noindent
{\bf Acknowledgements} 

This work was supported in
part by the U.S. Department of Energy Grant Numbers
DE-FG03-98ER41076 and DE-FG02-01ER45684.

\end{document}